\documentclass[aps,prx,superscriptaddress,nopacs,amsmath,amssymb,letter,reprint]{revtex4-1}

\usepackage{color}
\usepackage{graphicx}
\usepackage{dcolumn}
\usepackage{bm}

\begin{document}

\title{Hierarchical spin-orbital polarisation of a giant Rashba system} 

\author{L.~Bawden}
\affiliation {SUPA, School of Physics and Astronomy, University of
St. Andrews, St. Andrews, Fife KY16 9SS, United Kingdom}

\author{J.~M.~Riley}
\affiliation {SUPA, School of Physics and Astronomy, University of
St. Andrews, St. Andrews, Fife KY16 9SS, United Kingdom}
\affiliation{Diamond Light Source, Harwell Campus, Didcot, OX11 0DE, United Kingdom}

\author{C.~H. Kim}
\affiliation{Department of Applied Physics, Cornell University, Ithaca, New York 14853, USA}

\author{R.~Sankar}
\affiliation{Center for Condensed Matter Sciences, National Taiwan University, Taipei 10617, Taiwan}
\affiliation{Institute of Physics, Academia Sinica, Taipei 11529, Taiwan}

\author{E.~J.~Monkman}
\author{D.~E.~Shai}
\author{H.~I.~Wei}
\author{E.~Lochocki}
\affiliation{Laboratory of Atomic and Solid State Physics, Department of Physics, Cornell University, Ithaca, New York 14853, USA}

\author{J.~W.~Wells}
\affiliation{Department of Physics, Norwegian University of Science and Technology (NTNU), N-7491 Trondheim, Norway}

\author{W. Meevasana}
\affiliation {School of Physics, Suranaree University of Technology, Nakhon Ratchasima, 30000, Thailand} 
\affiliation{NANOTEC-SUT Center of Excellence on Advanced Functional Nanomaterials, Suranaree University of Technology, Nakhon Ratchasima 30000, Thailand}

\author{T.~K.~Kim}
\author{M.~Hoesch}
\affiliation{Diamond Light Source, Harwell Campus, Didcot, OX11 0DE, United Kingdom}

\author{Y.~Ohtsubo}
\author{P.~Le~F{\`e}vre}
\affiliation{Synchrotron SOLEIL, CNRS-CEA, L'Orme des Merisiers, Saint-Aubin-BP48, 91192 Gif-sur-Yvette, France}

\author{C.~J. Fennie}
\affiliation{Department of Applied Physics, Cornell University, Ithaca, New York 14853, USA}

\author{K.~M.~Shen}
\affiliation{Laboratory of Atomic and Solid State Physics, Department of Physics, Cornell University, Ithaca, New York 14853, USA}
\affiliation{Kavli Institute at Cornell for Nanoscale Science, Ithaca, New York 14853, USA}

\author{F.~C.~Chou}
\affiliation{Center for Condensed Matter Sciences, National Taiwan University, Taipei 10617, Taiwan}

\author{P.~D.~C.~King}
\altaffiliation{Corresponding author: philip.king@st-andrews.ac.uk}
\affiliation {SUPA, School of Physics and Astronomy, University of
St. Andrews, St. Andrews, Fife KY16 9SS, United Kingdom}

\begin{abstract} The Rashba effect is one of the most striking manifestations of spin-orbit coupling in solids, and provides a cornerstone for the burgeoning field of semiconductor spintronics. It is typically assumed to manifest as a momentum-dependent splitting of a single initially spin-degenerate band into two branches with opposite spin polarisation. Here, combining polarisation-dependent and resonant angle-resolved photoemission measurements with density-functional theory calculations, we show that the two ``spin-split'' branches of the model giant Rashba system BiTeI additionally develop disparate orbital textures, each of which is coupled to a distinct spin configuration. This necessitates a re-interpretation of spin splitting in Rashba-like systems, and opens new possibilities for controlling spin polarisation through the orbital sector. 
\end{abstract}

\date{\today}
\maketitle

\section*{Introduction}
The ability to controllably engineer spin splittings of electronic states is a key goal in the search for spintronic materials~\cite{koo_control_2009}. A particularly successful strategy has been the lifting of spin degeneracy via spin-orbit coupling in the presence of structural inversion asymmetry. Termed the Rashba or Bychov-Rashba effect~\cite{bychkov_properties_1984} this manifests through a spin-momentum locking of the quasiparticles, stabilising a pair of Fermi surfaces which are typically assumed to exhibit counter-rotating chiral spin textures~\cite{lashell_spin_1996}. The ability to electrostatically control the strength of such spin splitting~\cite{nitta_gate_1997,grundler_large_2000,king_large_2011,bahramy_emergent_2012} has led to prospects for all-electrical manipulation of electron spin precession~\cite{datta_electronic_1990}, offering new prototypical schemes of semiconductor spintronics~\cite{koo_control_2009}. The quest to practically realise such devices, and to operate them without cryogenic cooling, has motivated a major search for materials which can host stronger spin splittings than can typically be achieved in conventional semiconductors~\cite{ast_giant_2007,king_large_2011,sakamoto_valley_2013,yuan_zeeman-type_2013,riley_direct_2014}.

A giant Rashba-like spin splitting was recently discovered for bulk conduction and valence band states of the bismuth tellurohalide semiconductors~\cite{ishizaka_giant_2011,bahramy_origin_2011,murakawa_detection_2013}. Arising due to a combination of bulk inversion asymmetry (Fig.~\ref{f:overview}(b)), strong atomic spin-orbit coupling, and a negative crystal field splitting of the valence bands~\cite{bahramy_origin_2011}, Rashba parameters amongst the highest of any known material have been uncovered, together with a counter-rotating chiral Fermi surface spin texture~\cite{ishizaka_giant_2011,landolt_disentanglement_2012}. Exploiting element- and orbital-selective angle-resolved photoemission (ARPES), we show here that a complex interplay between atomic, orbital, and spin degrees of freedom significantly enriches this picture. We expect our findings to be broadly applicable across other strong spin-orbit Rashba systems.

\section*{Results}
\begin{figure}
\begin{center}
\includegraphics[width=\columnwidth]{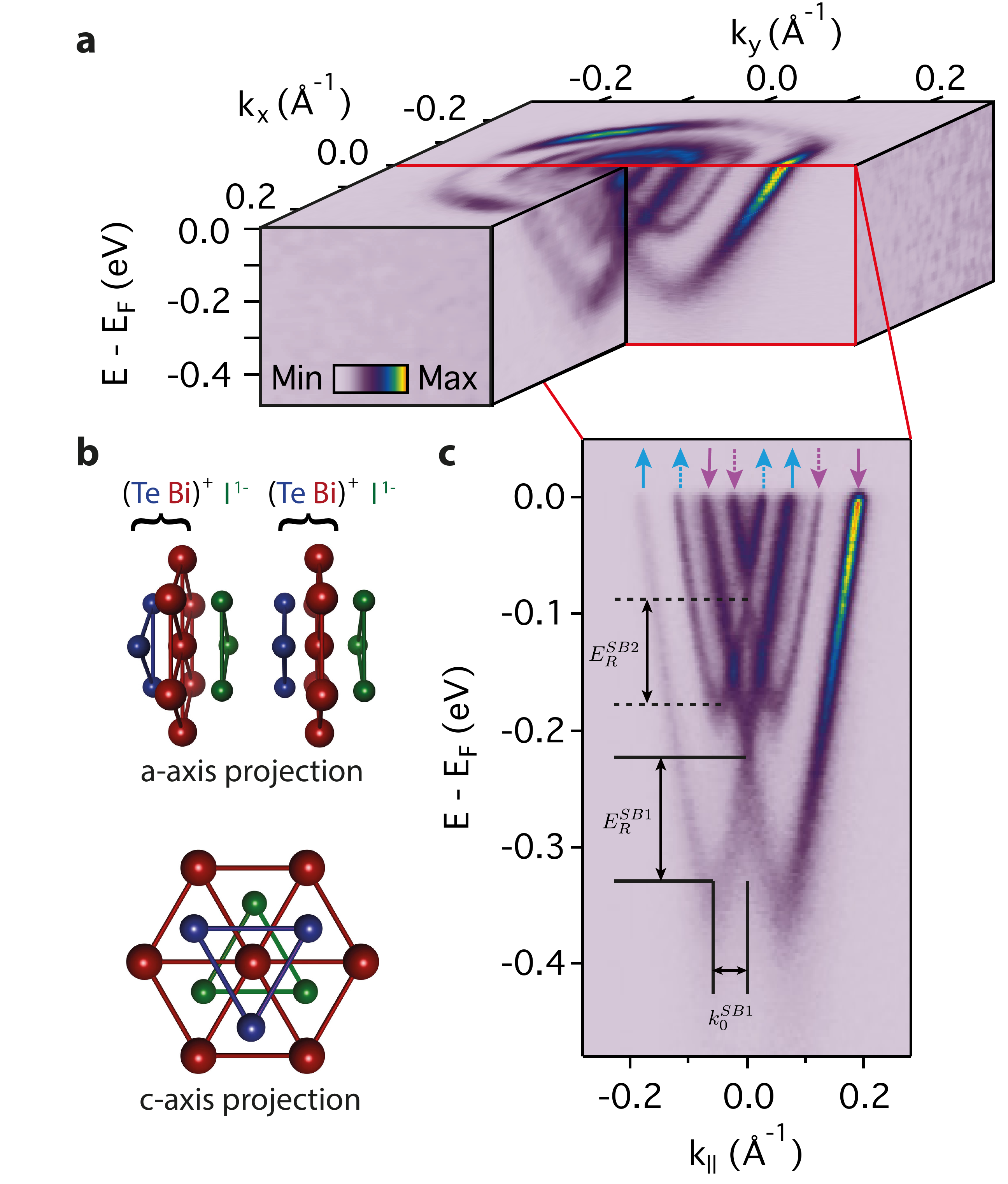}
\caption{ \label{f:overview} {\bf Surface electronic structure of BiTeI.} (a) ARPES measurements of the Fermi surface and near-$E_F$ band dispersions measured ($h\nu=52$~eV, $p$-polarisation) from the Te-terminated surface of BiTeI. A lack of inversion symmetry of the bulk crystal structure (b) together with strong spin-orbit coupling mediates a large Rashba-like spin splitting. Additionally, the polar nature of the Te-terminated surface induces a strong downward band bending, causing a ladder of Rashba-split subband (SB) states to emerge in the near-surface quantum well. (c) These are clearly resolved in measurements of the dispersion along $\Gamma-M$. The conventional spin texture associated with such Rashba splitting is shown schematically by coloured arrows, with the spin expected to lie predominantly in the surface plane. }
\end{center}
\end{figure}

First, we summarise the generic electronic structure of the Te-terminated surface of BiTeI (Fig.~\ref{f:overview}). Previous measurements have shown this termination to support a near-surface electron accumulation~\cite{ishizaka_giant_2011,crepaldi_giant_2012}. We clearly observe two 2D subbands (SB1 and SB2 in Fig.~\ref{f:overview}(c), see also Supplementary Fig.~S1) formed within the resulting quantum well. Each subband hosts two branches with a separation that grows approximately linearly with momentum away from their crossing at $k=0$. This is a hallmark of Rashba-like spin splitting. We extract a large Rashba energy $E_R = {120 \pm 10}$~meV (${85 \pm 5}$~meV) and momentum offset of the band bottom $k_0 = {0.055 \pm 0.005}$~\AA$^{-1}$ (${0.050 \pm 0.005}$~\AA$^{-1}$) for the first (second) subband, respectively, supporting previous studies which established this material as a model host of giant spin splittings~\cite{ishizaka_giant_2011,bahramy_origin_2011,landolt_disentanglement_2012}.

\begin{figure*}
\begin{center}
\includegraphics[width=\textwidth]{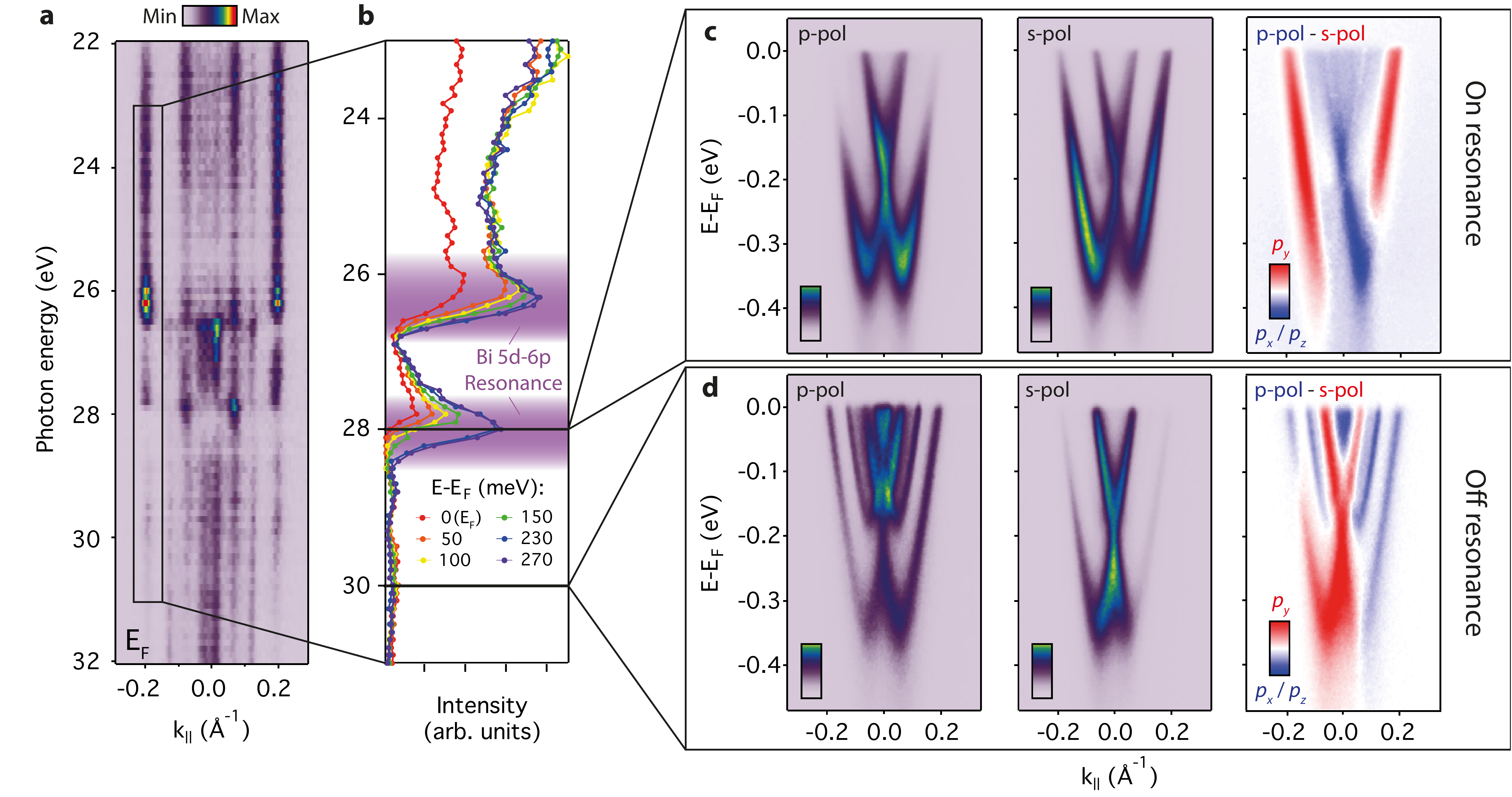}
\caption{ \label{f:disp} {\bf Disentangling intertwined atomic and orbital characters.} (a) Fermi level momentum distribution curve ($E_F\pm15$~meV) measured along $\Gamma-M$ as a function of probing photon energy using $p$-polarised light. No dispersion is observed, indicating two-dimensional states consistent with our assignment of quantum well subbands, while strong matrix element variations give rise to pronounced intensity modulations. These are shown for the outermost band ($k_F\approx-0.2$~\AA$^{-1}$) as a function of binding energy in (b), revealing characteristic intensity enhancement due to resonant photoemission at the Bi O-edge. (c,d) Corresponding dispersions (along $\Gamma-M$) measured on- ($h\nu=28$~eV) and off- ($h\nu=30$~eV) resonance, respectively, with $p$-polarised (left) and $s$-polarised (centre) light. The difference between these (right) indicates a band and elemental-dependent orbital polarisation.}
\end{center}
\end{figure*}
While typically treated in a single-band picture, the electronic wavefunction for each branch of a Rashba-split state can more generally be written as $\Psi=\sum_{i,\tau,\sigma}c_{i,\tau}^\sigma\psi_{i,\tau}$, where, following the notation in Ref.~\cite{zhu_layer-by-layer_2013}, $i$ is the atomic index, $\tau\in\{p_x,p_y,p_z\}$ and $\sigma$ are the orbital and spin indices, respectively, $\psi_{i,\tau}$ are atomic wavefunctions, and $c_{i,\tau}^\sigma$ are complex coefficients. Neglecting spin-orbit coupling, our calculations predict a conduction band in BiTeI predominantly derived from Bi~$p_z$ orbitals (see Supplementary Fig.~S2). Including such effects, however, not only permits it to become strongly spin-split via Rashba-like interactions, but also promotes significant orbital mixing. In general, therefore, multiple $c_{i,\tau}^\sigma$ can be expected to become non-negligible. For a complete description of the Rashba-split states, it is therefore essential to consider the interplay of the underlying atomic, orbital, and spin components. To disentangle these, we combine two powerful features of ARPES: characteristic selection rules for photoemission using linearly-polarised light, allowing us to directly probe the orbital wavefunction~\cite{damascelli_angle-resolved_2003,cao_mapping_2013}, and resonant photoemission to provide elemental sensitivity~\cite{hufner_photoelectron_1996}. Such resonant enhancements are evident in Fig.~\ref{f:disp}(a,b). They cause the spectral weight of the conduction band states to become strongly peaked at photon energies around 26 and 28~eV, close in energy to the binding energy of the Bi~$5d_{5/2,3/2}$ core levels, with functional forms that are well described by Fano lineshapes. This points to a significant Bi-derived atomic character of the lowest conduction band states, consistent with theoretical calculations~\cite{bahramy_emergence_2012}. We exploit this, selectively probing ``on-resonance'' to unveil the Bi-projected spectral function.

\begin{figure}
\begin{center}
\includegraphics[width=\columnwidth]{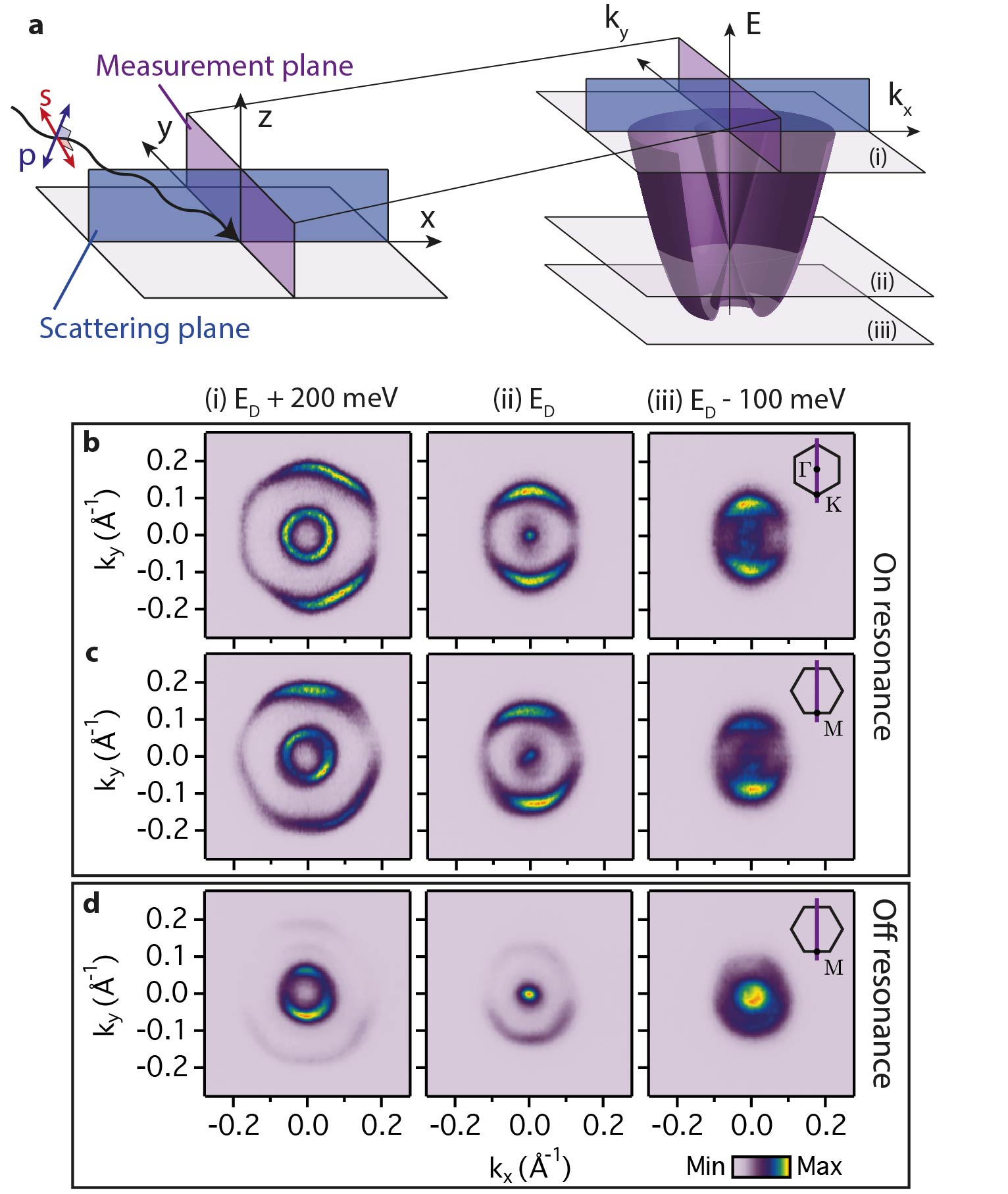}
\caption{ \label{f:cec} {\bf $p_y$-projected spectral weight distribution.} (a) Experimental geometry for our measurements, and (b-d) resulting constant binding energy surfaces measured using $s$-polarised light to probe the $p_y$ orbital character at 200~meV above, exactly at, and at 100~meV below the Dirac point ($E_D$) formed by the crossing of the two spin-split branches of the lowest subband. On-resonance measurements ($h\nu=28$~eV) with the scattering plane aligned to (b) $\Gamma-M$ and (c) $\Gamma-K$, and (d) off-resonance ($h\nu=30$~eV, scattering plane along $\Gamma-K$) measurements show pronounced angular variations in spectral weight, characteristic of strongly momentum-dependent orbital textures.}
\end{center}
\end{figure}

The resulting measurements of the dispersion reveal pronounced momentum-dependent spectral weight variations (Fig.~\ref{f:disp}(c)). We focus on the first subband (SB1), which is most clearly visible across our measurements. When probed using $p$-polarised light, we find stronger spectral weight for the inner branch of this subband, while the outer branch is significantly more pronounced when measured using $s$-polarisation. Selection rules~\cite{damascelli_angle-resolved_2003} dictate that, of the $p$-orbitals, the former measurement should be sensitive to $p_z$- and $p_x$-derived orbital character, while the transition matrix element is only non-vanishing for photoemission from $p_y$ orbitals in the latter case (our measurement geometry is shown in Fig.~\ref{f:cec}(a)). The asymmetric spectral weight distributions within and between these measurements immediately establishes that the two spin-split branches of the dispersion host a markedly different orbital makeup. Moreover, when measuring with a photon energy only 2~eV higher (Fig.~\ref{f:disp}(d)), we find an almost complete reversal of these matrix element variations, with greater spectral weight for the inner branch of the lowest subband when measuring using $s$-polarised light. No longer on resonance, these measurements will not {\it a priori} be dominated by the Bi-derived spectral weight, and we therefore conclude that the orbital textures of the Rashba-split states are also strongly dependent on their underlying atomic character, as discussed in detail below.

\begin{figure*}
\begin{center}
\includegraphics[width=\textwidth]{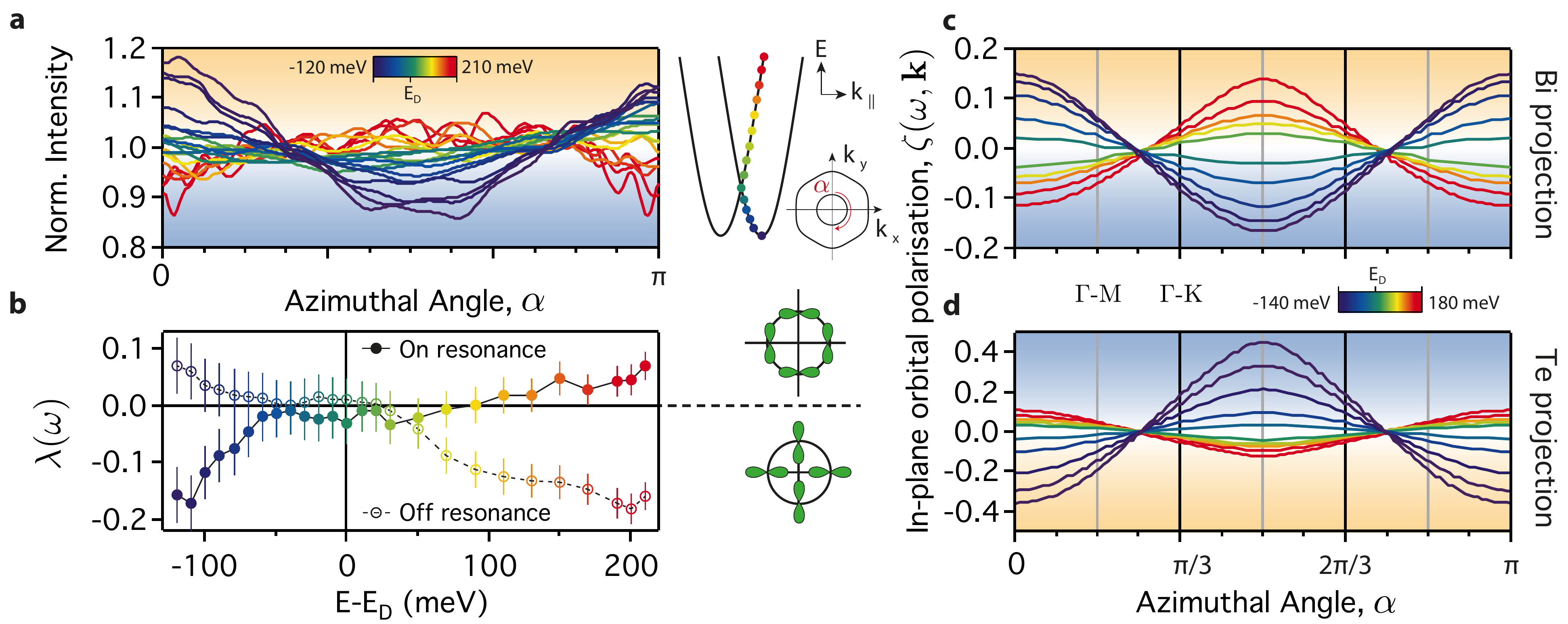}
\caption{ \label{f:orb} {\bf Mapping the angle-dependent orbital wavefunctions.} (a) Angular distribution of the Bi $p_y$-projected spectral weight distribution measured on-resonance around the inner band of CESs (see e.g. Fig.~\ref{f:cec}(b)), each normalised to its average value. Above (below) the Dirac point, $E_D$, this is peaked at an azimuthal angle of $\alpha=\pi/2$ (0 and $\pi$) indicative of a tangential (radial) in-plane orbital alignment. (b) The smooth evolution between these two configurations is captured by a relative spectral weight angular distribution factor (ADF) $\lambda(\omega)=[I_{\alpha=\pi/2}(\omega)-I_{\alpha=0,\pi}(\omega)]/[I_{\alpha=\pi/2}(\omega)+I_{\alpha=0,\pi}(\omega)]$ which crosses zero at the Dirac point within our experimental error. (c) This behaviour is fully captured by the momentum-dependent Bi $p_y:p_x$ in-plane orbital polarisation, $\zeta(\mathbf{\omega,k})$, extracted from our density-functional calculations. (d) The in-plane orbital polarisation is reversed for the Te-projected component, with radial alignment above the Dirac point and tangential alignment below $E_D$, as also captured by the opposite ADF measured off-resonance ($h\nu=30$~eV, see Fig.~\ref{f:cec}(c)), summarised in (b).}
\end{center}
\end{figure*}
Fig.~\ref{f:cec}(b,c) shows constant binding energy surfaces (CESs) measured on-resonance with $s$-polarised light, to selectively probe the momentum-space distribution of the Bi $p_y$ orbital character. The outer band of each CES exhibits strong spectral weight along our measurement direction (at positive and negative $k_y$), with a pronounced suppression at positive and negative $k_x$, over the entire occupied bandwidth of the conduction bands. This remains qualitatively unchanged under rotation of the azimuthal orientation of the sample. Such behaviour is indicative of a $p_y$-like orbital aligned along our measurement direction, irrespective of whether the $\Gamma-M$ or $\Gamma-K$ crystallographic direction is oriented to this~\cite{cao_mapping_2013,zhu_layer-by-layer_2013}. The in-plane Bi-projected charge density must therefore be oriented such that its lobes point predominantly radially outwards from the outer CES.

A similar conclusion can be drawn for the inner band of the CES at energies below the Dirac point formed by the crossing of the two spin-split branches of the dispersion. As already evident 100~meV below the Dirac point in Fig.~\ref{f:cec}(b,c), and more clearly shown in the angular dependence of spectral weight extracted around the CES in Fig.~\ref{f:orb}(a), the spectral weight is still peaked along the $k_y$ direction ($\alpha=0$ and $\pi$ in Fig.~\ref{f:orb}(a)), with a suppression of spectral weight along $k_x$ ($\alpha=\pi/2)$. Indeed, while the band edge turning point in the dispersion leads to a van Hove singularity in the density of states~\cite{ast_local_2007} the inner band of the CES below the Dirac point smoothly evolves into the outer band as it moves through this turning point, and thus hosts qualitatively the same orbital texture.

Moving up through the Dirac point, however, the angular distribution of spectral weight around the inner CES gradually flattens and then inverts to become peaked at $\alpha=\pi/2$. This points to a tangential alignment of the in-plane Bi-derived orbital character above the Dirac point, with stronger $p_y$ character at positive and negative $k_x$ than positive and negative $k_y$. This behaviour is well-reproduced by our first-principles calculations. As shown in Fig.~\ref{f:orb}(c), the in-plane $p_y:p_x$ orbital polarisation around the calculated CESs of the lower subband show a clear peak at $\alpha=\pi/2$ above the Dirac point, and 0 and $\pi$ below.
To capture this generic behaviour, we extract a relative spectral weight angular distribution factor from our measurements :
\begin{equation}\label{e:SWDF}
\lambda(\omega)=\frac{I_{\alpha=\pi/2}(\omega)-I_{\alpha=0,\pi}(\omega)}{I_{\alpha=\pi/2}(\omega)+I_{\alpha=0,\pi}(\omega)},
\end{equation}
where $I_\alpha(\omega)$ is the spectral weight at angle $\alpha$ and energy $\omega$. This provides an experimental measure of the calculated ``orbital polarisation ratio'' used to describe topological states in Ref.~\cite{cao_mapping_2013}. Here, this parameterises the relative strength of spectral weight, and thus $p_y$ orbital character, along the $k_x$ and $k_y$ directions of our measured CESs, reflecting the strength ($|\lambda|$) and alignment of radial ($\lambda<0$) and tangential ($\lambda>0$) in-plane orbital textures. As shown in Fig.~\ref{f:orb}(b), we find this is positive at energies above the Dirac point for the on-resonance measurements, consistent with the dominantly tangential in-plane Bi-projected charge density distribution of the inner band assigned above. With decreasing energy towards the Dirac point, $\lambda$ is suppressed to zero, indicating a loss of orbital polarisation, before growing again but with opposite sign below the Dirac point. Thus, we have experimentally observed a gradual crossover from a radial to a tangential alignment of the in-plane Bi orbital character with the switch occurring exactly at the Dirac point within our experimental error.  This is strikingly similar to a switch in orbital texture that was recently reported at the Dirac point of the spin-polarised surface states of topological insulators~\cite{cao_mapping_2013,zhu_layer-by-layer_2013}. Our observation of such a crossover in a topologically-trivial compound establishes this as a general feature of strongly spin-orbit coupled systems, a point we return to below.

Intriguingly, extracting $\lambda(\omega)$ from our measurements performed off-resonance, we find an opposite trend (Fig.~\ref{f:orb}(b)). Above the Dirac point, $\lambda<0$ indicating a radial alignment of the in-plane orbital-projected charge density, as also clearly evident in the spectral weight distribution visible in Fig.~\ref{f:cec}(d). This is again smoothly suppressed to zero approaching the Dirac point before becoming positive, revealing a tangential orbital configuration, at energies below Dirac point. We attribute this as a signature of the in-plane Te orbital polarisation, which well matches that found in our calculations (Fig.~\ref{f:orb}(d)). It can be seen in our measurements here at 30~eV photon energy as, with no Bi resonant enhancement, the photoemission cross-section is higher for Te~$5p$ than Bi~$6p$ orbitals~\cite{Yeh_atomic_1985} and, with Te being located right at the surface, there is no depth-dependent attenuation due to the surface sensitivity of photoemission. Our calculations (Supplementary Fig.~S2) reveal a strong in-plane Te character of the inner branch of the dispersion, which can thus dominate the spectral weight of the inner branch of the CES in these off-resonance measurements. For the outer branch, however, there is little calculated in-plane Te weight (Supplemental Fig.~S2), and so the weak spectral features visible for the outer branch in Fig.~\ref{f:cec}(d) is still reflective of the Bi in-plane orbital texture.

\section*{Discussion}
\begin{figure*}
\begin{center}
\includegraphics[width=\textwidth]{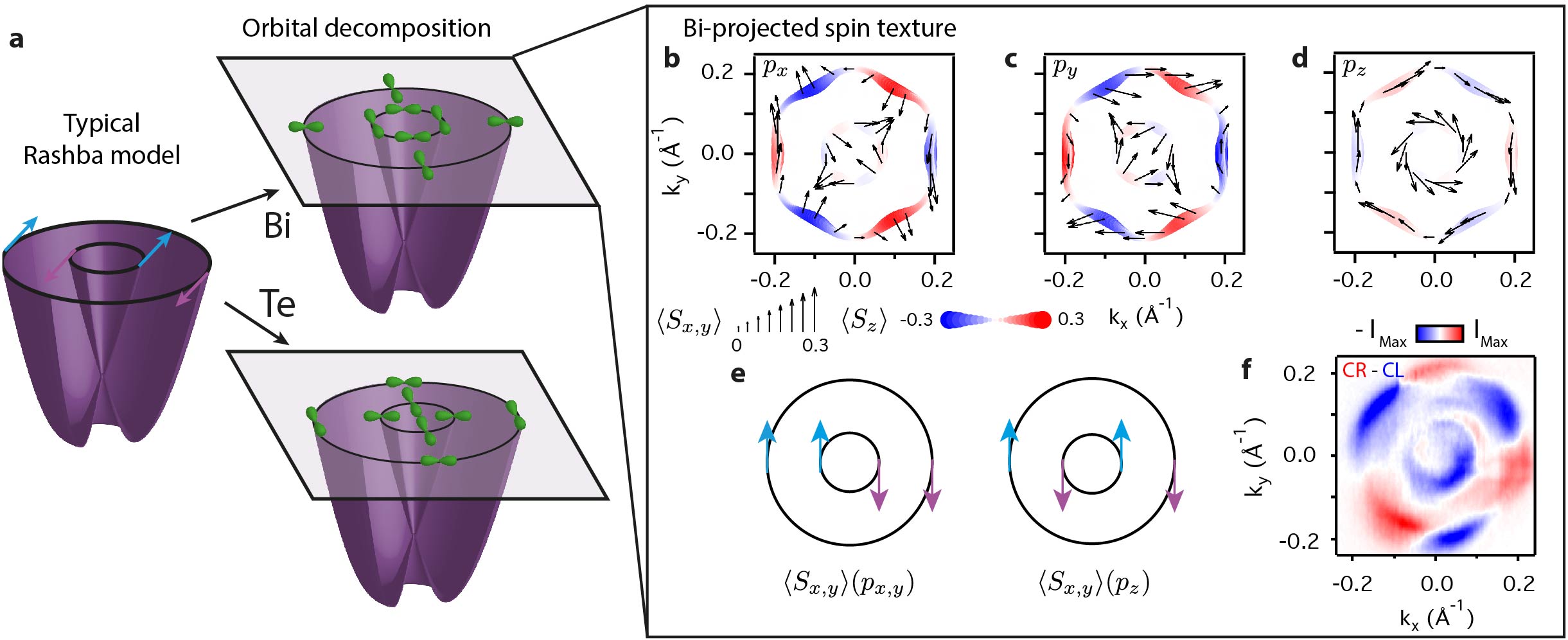}
\caption{ \label{f:CD} {\bf Hierarchy of atomic, spin, and orbital coupling in BiTeI.} (a) Schematic representation of disparate in-plane orbital textures uncovered for the inner and outer ``spin-split'' branches a model Rashba system. Corresponding spin textures calculated from density-functional theory for Bi (b) $p_x$, (c) $p_y$, and (d) $p_z$ projections of CESs 200~meV above the Dirac point. The in-plane spin texture is shown by arrows and the out-of-plane by the background colour, both in units of $\hbar/2$. (e) Schematic of the in-plane spin textures coupled to the net in-plane and out-of-plane orbital textures. (f) Circular dichroism measurements performed on-resonance ($h\nu=28$~eV, CES at $E_D+200$~meV), revealing a significant six-fold modulation for the outer band indicative of pronounced out-of-plane spin canting.}
\end{center}
\end{figure*}

Together, these measurements and calculations reveal that spin-orbit coupling induces a complex atomic- and momentum-dependent hierarchy of orbitally-polarised components of the underlying electronic structure in BiTeI, summarised schematically in Fig.~\ref{f:CD}. Our calculations additionally reveal how each orbital component is, in turn, coupled to a disparate spin texture. We illustrate this for the Bi-derived states in Fig.~\ref{f:CD}(b-d); the Te-derived component is additionally shown in Supplementary Fig.~S2. The in-plane spin-texture $\langle{S_{x,y}}\rangle$ projected onto Bi $p_z$ orbitals yields a conventional counter-rotating chiral spin texture of neighbouring CESs at energies above the Dirac point, characteristic of classic Rashba systems~\cite{bychkov_properties_1984,lashell_spin_1996} and indeed observed experimentally for BiTeI~\cite{ishizaka_giant_2011,landolt_disentanglement_2012}. In contrast, the spin texture is significantly more complex for the $p_x$ and $p_y$ orbital projection, with the in-plane spin component switching between tangential and radial around the CES. This results from a coupling of the spin to the characteristic orbital textures, and is similar to that found recently for topological surface states~\cite{zhang_spin-orbital_2013,cao_coupled_2012,zhu_photoelectron_2014}. 

For example, combined with our uncovered tangential (radial) in-plane Bi orbital texture above (below) the DP, this leads to a net clockwise spin rotation for the in-plane orbital projection of {\it both} the inner and outer CESs. This is in strong departure from the conventional picture of Rashba-like spin polarisation. Such in-plane orbital textures likely also underlie the unconventional spin topology predicted for certain $p_{x,y}$-derived states in Bi/Cu(111)~\cite{mirhosseini_unconventional_2009} and Pb/Cu(111)~\cite{bihlmayer_enhanced_2007} surface alloys. They are also broadly consistent with recent first-principles calculations that suggest the Rashba parameter can be orbital-dependent in bismuth tellurohalides~\cite{zhu_orbital-dependent_2013}, although here we reveal how orbital mixing can lock the spin components on the nominally ``spin-split'' CESs together, stabilising pronounced components of the underlying wavefunction that host markedly non-Rashba-like spin textures in this system. Similar considerations hold for the Te-derived components, although with additional variations in the magnitude of the spin components projected onto out-of-plane vs. in-plane orbital components between the CESs due to a greater out vs. in-plane orbital polarisation for Te (Supplementary Figs.~S2 and S3). With each orbital component locked to a different spin texture, the fundamental requirement from time-reversal symmetry of spin-degeneracy at the Kramer's point (the Dirac point formed in this system at $k=0$) ensures equal contribution of in-plane $p_x$ and $p_y$ orbitals at this point. Thus, the vanishing of orbital polarisation at the Dirac point observed experimentally above can be simply viewed as an orbital analogue of Kramer's spin degeneracy in time-reversal symmetric systems.

Away from the Dirac point, our calculations additionally predict a strong canting of the spin out of the surface plane for the Bi-derived orbitals. This grows in magnitude with increasing energy away from the Dirac point, where the CESs become increasingly hexagonally warped~\cite{ishizaka_giant_2011}. We show evidence for this through circular dichroism measurements. We attribute such dichroism as a signature of unquenched orbital angular momentum~\cite{park_orbital-angular-momentum_2011}, which our calculations reveal is large and locked approximately opposite to the spin angular momentum due to the strong spin-orbit coupling. Circular dichroism is known to show a complex dependence on photon energy in this system~\cite{crepaldi_momentum_2014} as well as in other layered compounds such as Bi$_2$Te$_3$ (Ref.~\onlinecite{HasanCD}), which can naturally arise as a consequence of inter-layer photoelectron interference~\cite{crepaldi_momentum_2014,zhu_layer-by-layer_2013,zhu_photoelectron_2014}. To simplify such effects, we again perform our measurements on-resonance, selectively enhancing states of Bi character (Fig.~\ref{f:CD}(f)). We find that the outer (hexagonally-warped) band in such measurements develops a pronounced six-fold modulation, which has previously been shown to reflect out-of-plane spin-canting in topological insulators~\cite{wang_observation_2011,jung_warping_2011,bahramy_emergent_2012}. Here, this provides the first experimental evidence for deviations from simple in-plane chiral spin textures in BiTeI.

As evident in Fig.~\ref{f:CD}(b-d), $\langle{S_z}\rangle$ is substantially larger when projected onto the in-plane orbital components than for the $p_z$ projection. The emergence of a large out-of-plane spin canting in this system is thus intricately tied to the development of the in-plane orbital texture away from the Dirac point observed here. Together, our measurements and calculations establish a powerful role of in-plane atomic orbitals shaping the spin structure of the model Rashba compound BiTeI, revealing a complex interplay between atomic makeup, anisotropic orbital textures, and spin-momentum locking. While small quantitative variations may occur due to near-surface potential contributions to the Rashba spin splitting and possible surface-induced orbital reconstructions here, our findings should be broadly applicable to the spin-split bulk electronic structure of BiTeI as well as being generic to other strong spin-orbit Rashba systems, suggesting new routes to control spin splitting through the orbital sector. For example, exploiting structure-property relations to tune the competition of atomic spin-orbit coupling and crystal field splitting will allow control over the ratio of in-plane and out-of-plane orbital polarisation, thus modulating the degree of non-Rashba-like spin components and out-of-plane spin canting. Together, this promises new prospects for the targeted design of optimised spintronic materials.

\section*{Materials and Methods}

{\bf Angle-resolved photoemission:} ARPES measurements were performed at the CASSIOPEE beamline of SOLEIL synchrotron (France) and I05 beamline of Diamond Light Source (UK). Single-crystal samples of BiTeI, grown by chemical vapour transport, were cleaved {\it in-situ} at the measurement temperature of 10~K. Measurements were performed using linear-horizontal, linear-vertical, and circularly-polarised synchrotron light at the photon energies described in the text, and reproduced on multiple samples. Scienta R4000 hemispherical electron analysers were employed, with a vertical entrance slit and the light incident in the horizontal plane as shown in Fig.~\ref{f:cec}(a). BiTeI is known to have domains of mixed Te and I termination, supporting surface electron and hole accumulation layers, respectively~\cite{crepaldi_giant_2012}. By monitoring the relative spectral weight of electron- and hole-like bands crossing the Fermi level and from characteristic core-level shifts in x-ray photoemission spectra, we aligned the synchrotron light spot on Te-terminated domains of the sample. From core-level spectra, we estimate an upper limit of 2\% I-terminated regions within our probing region for the data shown here. 

{\bf Density-functional theory calculations:} Density-functional theory (DFT) was performed within the generalized gradient approximation (GGA). We used the DFT code, OpenMX \cite{openmx}, based on the linear-combination-of-pseudo-atomic-orbitals (LCPAO) method \cite{ozaki}.
Spin-orbit interaction was included via the norm-conserving, fully relativistic $j$-dependent pseudopotential scheme in the non-collinear DFT formalism \cite{theurich}. We model the Te-terminated BiTeI surface electronic structure via a supercell calculation including a slab of 60 atomic layers and a vacuum region of $\sim\!20$~\AA{} thickness. To calculate the spin and orbital angular momentum for specific $k$-points, we used the LCPAO coefficients of local atoms.

\section*{Acknowledgements}

We acknowledge M.S. Bahramy for useful discussions. This work was supported by the Engineering and Physical Sciences Research Council, UK (Grant Nos.~EP/I031014/1, EP/M023427/1 and EP/G03673X/1), the Ministry of Science and Technology in Taiwan (project number MOST-102-2119-M-002 -004), TRF-SUT Grant RSA5680052, the National Science Foundation (Grant Nos. DMR-0847385 and DMR-1120296), and the Office of Naval Research (Grant No. N00014-12-1-0791). PDCK acknowledges support from the Royal Society through a University Research Fellowship. HIW acknowledges support from the NSF IGERT program (DGE-0903653) and the National Science Foundation Graduate Research Fellowship under Grant no. DGE-1144153. We acknowledge SOLEIL for provision of synchrotron radiation facilities (CASSIOPEE beamlime) and also thank Diamond Light Source for access to beamline I05 (proposal number SI9427) that contributed to the results presented here.

\section*{Competing Interests} The authors declare that they have no competing interests.

\section*{Supplemental Materials}
\noindent Fig. S1: Photon energy-dependent ARPES\\
Fig. S2: Calculated surface electronic structure of BiTeI\\
Fig. S3: Coupled Te spin-orbital texture\\

\



\renewcommand{\thefigure}{S\arabic{figure}}   
\setcounter{figure}{0}

\

\begin{figure*}
\begin{center}
\includegraphics[width=0.85\textwidth]{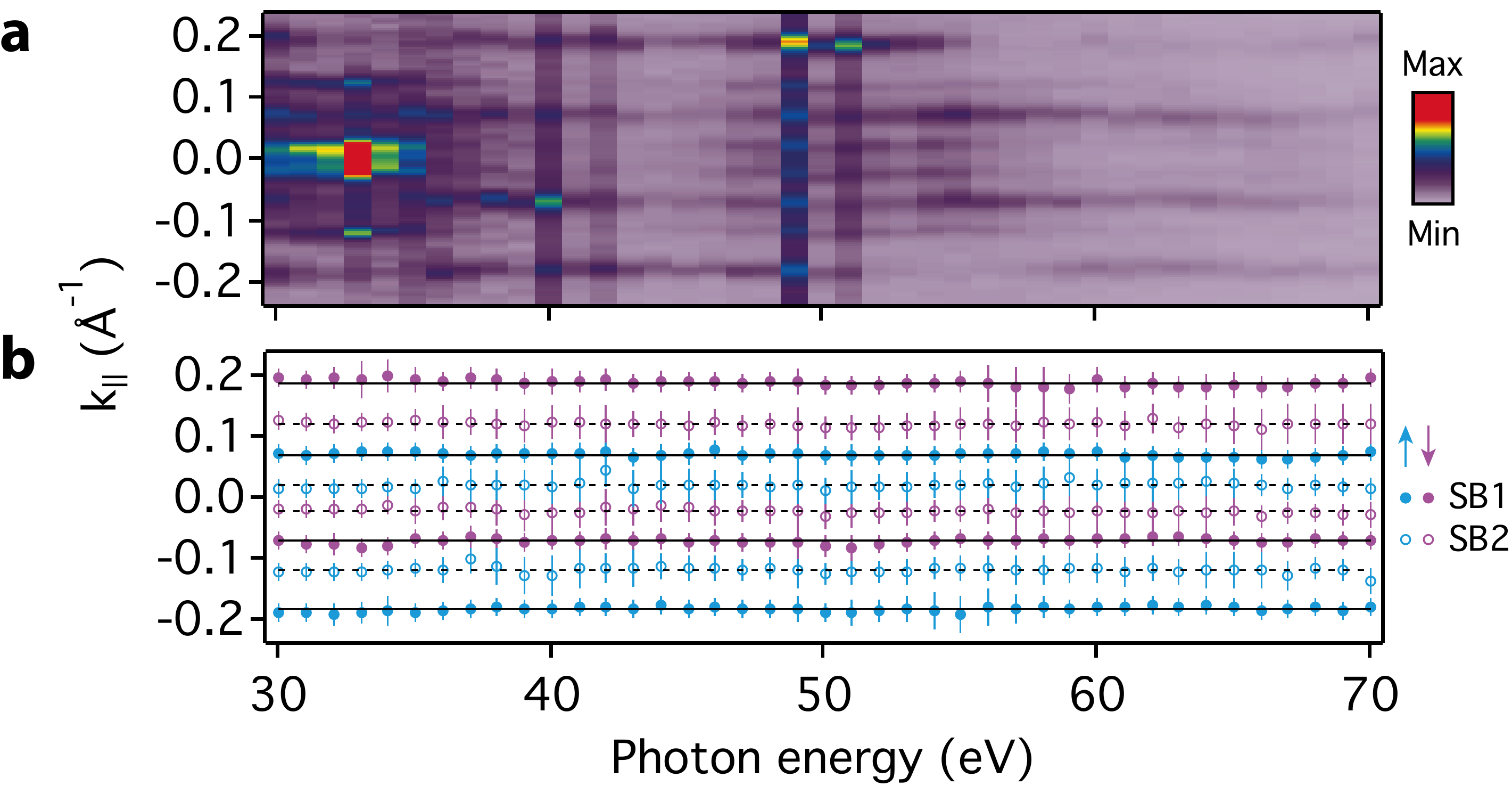}
\caption{ \label{f:s0} {\bf Photon energy-dependent ARPES.} (a) MDC at the Fermi level ($E_F\pm15$~meV) measured along the $\Gamma-M$ direction as a function of photon energy using $p$-polarised light. This shows non-dispersive features indicative of two-dimensional electronic states. (b) This is confirmed by the Fermi momenta extracted by fitting MDCs. The straight lines show the average extracted $k_F$ for each band. We thus assign all eight Fermi crossings as arising from two-dimensional subband states formed within a near-surface quantum well arising from the downward band bending potential at the Te-terminated surface. This is consistent with their spectral weight distribution: the states arising from the second subband (SB2), being less strongly bound than SB1 within the almost-triangular quantum well potential at the surface, will have an envelope wavefunction with a larger spatial extent perpendicular to the sample surface~[39]. This would be expected to give rise to a periodic variation of the photoemission spectral weight with photon energy peaked over a narrower range than for SB1~[40], as observed experimentally. }
\end{center}
\end{figure*}

\begin{figure*}
\begin{center}
\includegraphics[width=0.85\textwidth]{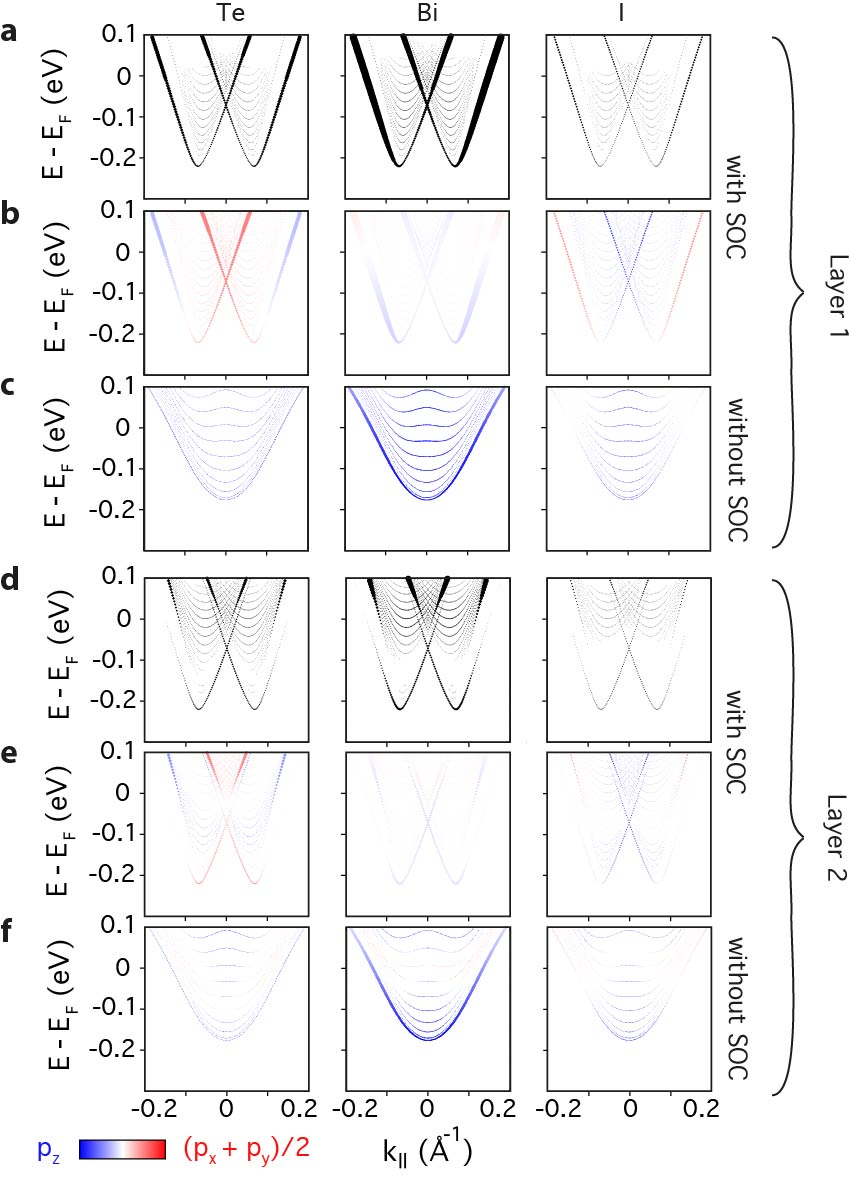}
\caption{ \label{f:s1} {\bf Calculated surface electronic structure of BiTeI.} (a,d) Atomic weight and (b,e) ratio of out-of-plane to average in-plane orbital makeup of the surface electronic structure of BiTeI, from Te-terminated supercell calculations within density-functional theory including spin-orbit coupling. (c,f) Equivalent orbital-projected electronic structure from calculations neglecting spin-orbit coupling. Projections onto the (a-c) first and (d-f) second Te-Bi-I trilayers of the supercell are shown.}
\end{center}
\end{figure*}

\begin{figure*}
\begin{center}
\includegraphics[width=\textwidth]{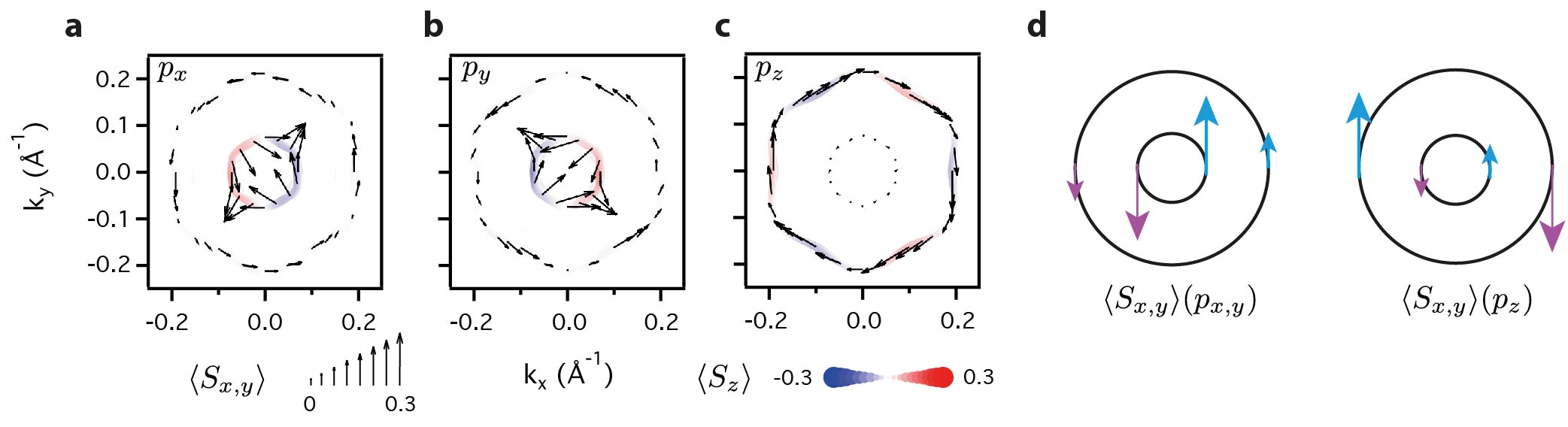}
\caption{ \label{f:s3} {\bf Coupled Te spin-orbital texture.} Spin textures for Te (a) $p_x$, (b) $p_y$, and (c) $p_z$ projections of CESs at 200~meV above the Dirac point. The in-plane spin texture is shown by arrows, the out-of-plane by the background colour, both in units of $\hbar/2$. (d) Schematic of the in-plane spin textures coupled to the net in-plane and out-of-plane orbital textures.}
\end{center}
\end{figure*}

\end{document}